\documentclass[conference]{IEEEtran}
\IEEEoverridecommandlockouts

\usepackage{cite}
\usepackage{amsmath,amssymb,amsfonts}
\usepackage{algorithmic}
\usepackage{graphicx}
\usepackage{textcomp}
\usepackage{bmpsize}
\usepackage{xcolor}
\usepackage{subcaption}
\usepackage{multirow}  
\usepackage{booktabs}  
\usepackage{listings}
\usepackage{adjustbox}
\usepackage{xspace}
\usepackage{dsfont}

\DeclareMathOperator*{\argmin}{arg\,min}

\usepackage{tikz} 
\usetikzlibrary{arrows.meta, positioning, shapes.geometric, calc, shapes.multipart, backgrounds, decorations.markings, fit} 

\usepackage[colorlinks=true,urlcolor=black]{hyperref}
\def\BibTeX{{\rm B\kern-.05em{\sc i\kern-.025em b}\kern-.08em
    T\kern-.1667em\lower.7ex\hbox{E}\kern-.125emX}}

\definecolor{customgray}{gray}{0.7} 
\definecolor{backcolour}{rgb}{0.95,0.95,0.92}

\lstdefinestyle{coloredlisting}{
    backgroundcolor=\color{backcolour},
    rulecolor=\color{backcolour!40!black},
    frameround=tttt,
    frame=single,
    basicstyle=\ttfamily\scriptsize,
    basewidth=0.50em,
    keywordstyle=\bfseries,
    xleftmargin=.02\linewidth,
    xrightmargin=.02\linewidth,
    escapeinside={<@}{@>},
    numbers=none,
    captionpos=b,
    breaklines=true  
}

\newcommand{\tool}{SALM\xspace}

\begin{document}

\title{Improving Generalization on Cybersecurity Tasks with Multi-Modal Contrastive Learning}

\author{
\IEEEauthorblockN{
Jianan Huang,
Rodolfo V. Valentim,\\
Luca Vassio, 
Matteo Boffa,
Marco Mellia
}
\IEEEauthorblockA{
Politecnico di Torino, Italy \\
\{first.second\}@polito.it
}
\and
\IEEEauthorblockN{Idilio Drago}
\IEEEauthorblockA{
Università di Torino, Italy \\
idilio.drago@unito.it
}
\and
\IEEEauthorblockN{
Dario Rossi
}
\IEEEauthorblockA{
Huawei Paris Research Center, France \\
dario.rossi@huawei.com
}
}

\maketitle

\begin{abstract}
The use of ML in cybersecurity has long been impaired by generalization issues: Models that work well in controlled scenarios fail to maintain performance in production. The root cause often lies in ML algorithms learning superficial patterns -- \emph{shortcuts} -- rather than underlying cybersecurity concepts. We investigate contrastive multi-modal learning as a first step towards improving ML performance in cybersecurity tasks. We aim at transferring knowledge from data-rich modalities, such as text, to data-scarce modalities, such as payloads.
We set up a case study on \textit{threat classification} and propose a two-stage multi-modal contrastive learning framework that uses textual vulnerability descriptions to guide payload classification. First, we construct a semantically meaningful embedding space using contrastive learning on descriptions. Then, we align payloads to this space, transferring knowledge from text to payloads. We evaluate the approach on a large-scale private dataset and a synthetic benchmark built from public CVE descriptions and LLM-generated payloads. The methodology appears to reduce shortcut learning over baselines on both benchmarks. We release our synthetic benchmark and source code as open source\footnote{\url{https://github.com/SmartData-Polito/LLM_semantic_alignment}}.
\end{abstract}

\begin{IEEEkeywords}
Contrastive Learning, Threat Classification.
\end{IEEEkeywords}

\section{Introduction}
\label{sec:intro}

Researchers across multiple disciplines have identified a growing ``credibility crisis'' in Machine Learning (ML), stemming from poor generalization~\cite{apruzzese2023sok, willinger2025something}: results that look promising during development often vanish once models are deployed. In cybersecurity, this problem is particularly clear. Models trained on limited proprietary datasets tend to learn superficial correlations (\textit{shortcuts}~\cite{geirhos2020shortcut}) rather than underlying cybersecurity semantics~\cite{emperor_has_no_clothes}. Arp et al.~\cite{arp2022and} identified ten pitfalls that inflate reported performance, while Pendlebury et al.~\cite{pendlebury2019tesseract} showed that temporal and spatial bias systematically degrade generalization in malware classification. Recent work confirms these limitations across security domains~\cite{risse2024uncovering, yang2021cade, zhao2025sweet}. Building models that can \emph{reason beyond pattern matching} is essential for cyber defense, where systems must contend with unpredictable zero-day threats~\cite{han2023anomaly}.

\begin{figure}[!t]
    \centering
    \includegraphics[width=0.9\linewidth]{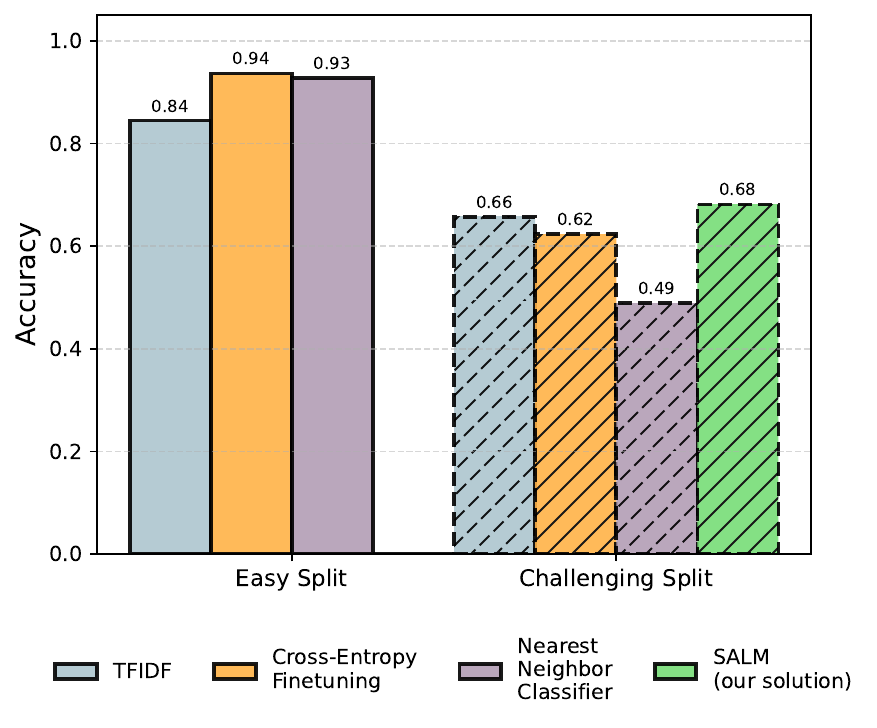}
    \caption{\textbf{High apparent accuracy masks poor generalization in threat classification.} While some models exceed 0.92 accuracy in the \textit{easy scenario} (solid bars), performance drops by almost 30\% in the \textit{challenging scenario} (dashed bars). \tool achieves the highest score in the challenging scenario.}
    \label{fig:intro_baseline_comparison}
\end{figure}

We make this problem concrete through \textit{threat classification}: the task is classifying malicious HTTP payloads into 15 attack types (e.g., SQL Injection, XSS). Figure~\ref{fig:intro_baseline_comparison} compares traditional approaches across two evaluation setups. In the \textit{easy scenario} (random split), all models achieve high accuracy (above 0.84). In the \textit{challenging scenario} (time-based split requiring generalization to temporally shifted data), every model experiences a sharp performance drop, illustrating the credibility crisis in cybersecurity ML.

We here give a first step to test an approach proven effective in computer vision~\cite{clip} and code understanding~\cite{wang_codet5}: true generalization arises when models leverage knowledge from data-rich modalities (e.g., text) and transfer it to data-scarce domains. In cybersecurity, abundant textual artifacts -- vulnerability reports, documentation, attack taxonomies -- form part of LLM pretraining corpora, and general-purpose LLMs have been shown to answer high percentages of cybersecurity questions~\cite{wang2025digital}. This motivates fully exploiting such knowledge for specialized tasks like payload classification, following the cross-modal alignment paradigm that has succeeded in vision-language models~\cite{clip, zhai2022lit}.

We propose \tool (Semantically Aligned Language Models), a two-stage contrastive learning framework. In Stage~1, we realign the LLM embedding space to cybersecurity through contrastive learning on vulnerability description triplets. In Stage~2, we transfer this knowledge to payloads via teacher-student alignment with a frozen text encoder. We evaluate \tool on a large-scale private corpus with temporal splits and a synthetic benchmark built from public CVE descriptions and LLM-generated payloads. \tool reaches 0.68 accuracy in the challenging scenario -- improving over Cross-Entropy Finetuning (0.62) and Nearest Neighbor (0.49) -- with similar gains also on the synthetic benchmark. 
Although significant room for improvement remains, these results suggest that contrastive multi-modal learning can help reduce shortcut learning. We release our synthetic benchmark and source code as open source to foster further studies on the approach.

\section{Background}\label{sec:background}

\subsection{Pretrained Models as Knowledge Encoders}

Modern Language Models (LMs) learn rich representations through self-supervised training on large text corpora. A widely used objective is \textit{causal language modeling}, where the model predicts the next token given its context, learning a mapping from input tokens to embeddings $e_i \in \mathbb{R}^d$. These token-level representations can be aggregated into sentence-level embeddings~\cite{Su2023Instructor} (e.g., via [CLS] token or mean pooling), producing compact representations where semantically similar concepts are located nearby in the vector space. This property makes pretrained LMs natural foundations for tasks requiring semantic reasoning -- including cybersecurity, where LLMs have been shown to encode substantial domain knowledge~\cite{wang2025digital}.

\subsection{Contrastive Learning, Triplet Loss, Cross-Modal Transfer}

Contrastive learning organizes an embedding space by pulling similar samples together while pushing dissimilar ones apart~\cite{khosla2020supervised}. The triplet loss~\cite{Schroff2015FaceNet} operates on triplets $(x^a, x^p, x^n)$ of anchor, positive (same class), and negative (different class) samples:
\begin{equation}
\mathcal{L}_{\text{triplet}} = \sum_{i=1}^{N} \max(0, d_{ap}^i - d_{an}^i + \alpha)
\end{equation}
where $d_{ap}^i$ and $d_{an}^i$ are anchor-positive and anchor-negative distances, and $\alpha$ is a margin. Only triplets violating the margin constraint contribute to the loss.

Cross-modal transfer enables a student model on a data-scarce modality to benefit from a teacher trained on richer data~\cite{zhai2022lit}. Given a frozen teacher encoder $f_t$ (e.g., for text) and a trainable student encoder $f_s$ (e.g., for payloads), modal alignment minimizes:
\begin{equation}
\mathcal{L}_{\text{align}} = \|f_s(x_s) - f_t(x_t)\|^2
\end{equation}
Freezing the teacher prevents catastrophic forgetting and provides stable target embeddings~\cite{zhai2022lit}, transferring the teacher's semantic geometry to the student.

\section{\tool: Semantically Aligned Language Models}
\label{sec:method}

Vulnerability descriptions provide high-level semantic structure: they explain \emph{what} an attack exploits and \emph{how} it operates, using terminology that persists across time. Network payloads, in contrast, contain noisy, low-level attack realizations where the same vulnerability manifests through countless variants and encoding schemes. This asymmetry drives the poor generalization observed in Figure~\ref{fig:intro_baseline_comparison}.

\begin{figure}[t]
    \centering
    \includegraphics[width=1\linewidth]{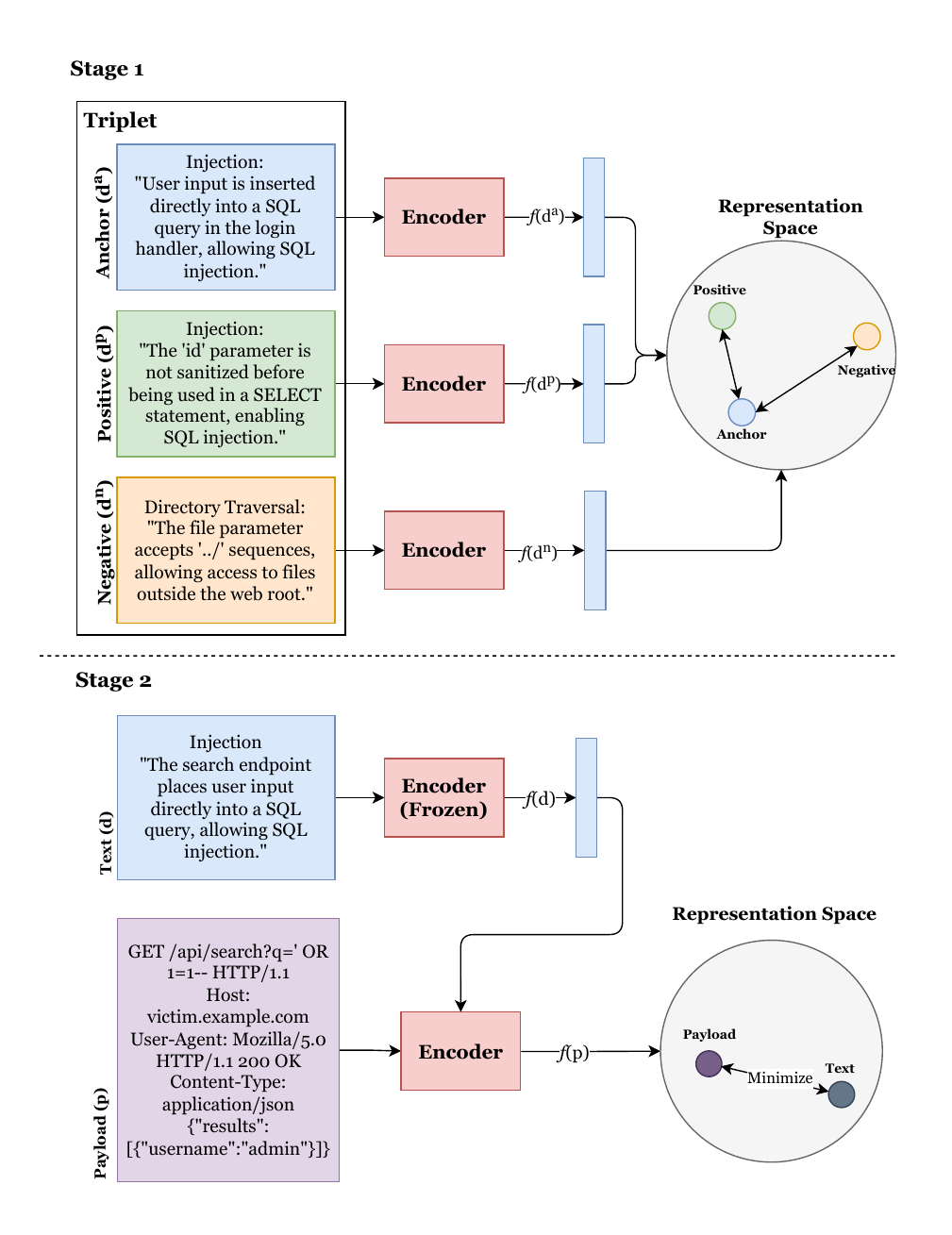}
    \caption{Two-stage training procedure of \tool. \textbf{Top}: Contrastive learning structures the text embedding space around vulnerability types. \textbf{Bottom}: Modal alignment transfers this structure to payloads via a frozen text encoder, producing a unified semantic space.}
    \label{fig:training_stages}
\end{figure}

We address this with \tool, a two-stage framework that transfers semantic structure from text to payloads. Stage~1 constructs a vulnerability-aware embedding space using contrastive learning on descriptions. Stage~2 aligns payload representations to this space through cross-modal alignment with a frozen text encoder. The result is a unified embedding space organized by vulnerability type, enabling semantic retrieval at inference. Figure~\ref{fig:training_stages} illustrates the process.

\subsection{Stage 1: Vulnerability-Aware Embedding Space}

The first stage learns a structured representation space from textual vulnerability descriptions using contrastive learning. For each vulnerability type ($y \in \{1, \ldots, 15\}$ in our case study), we sample triplets $(d^a, d^p, d^n)$ where anchor and positive share the same type and the negative differs. We train a text encoder $f_t: \mathcal{X}_t \rightarrow \mathbb{R}^{768}$ to minimize the triplet loss (Equation~1), enforcing tight intra-type clusters and clear inter-type separation. We initialize $f_t$ from \texttt{hkunlp/instructor-base}~\cite{Su2023Instructor}, a 110M-parameter T5-based model for instruction-aware embeddings.

This stage addresses a fundamental question: why not rely directly on pretrained LLM knowledge? While LLMs possess broad cybersecurity knowledge~\cite{wang2025digital}, contrastive training optimizes the embedding space to prioritize vulnerability-type boundaries specifically. Section~\ref{sec:results} provides visual evidence of this restructuring.

\subsection{Stage 2: Transferring Structure to Payloads}

Once the text encoder has learned a structured space, we transfer this structure to payloads. From the training split we build pairs $(d, p)$ where payload $p$ is linked to description $d$. We freeze $f_t$ from Stage~1 and train a payload encoder $f_p: \mathcal{X}_p \rightarrow \mathbb{R}^{768}$ (also initialized from \texttt{instructor-base}) to minimize the alignment loss (Equation~2):
\begin{equation}
\mathcal{L}_{\text{align}} = \frac{1}{M} \sum_{i=1}^{M} \|f_p(p_i) - f_t(d_i)\|^2
\end{equation}
Freezing $f_t$ prevents catastrophic forgetting of Stage~1 structure and provides stable target embeddings. After Stage~2, both encoders map their inputs into the same 768-dimensional space.

\subsection{Inference via Semantic Retrieval}

At test time, classification is performed through semantic retrieval. For each vulnerability type, we construct a prototype by encoding a short generic textual label (e.g., ``SQL injection attack'') with $f_t$: $\bar{e}_y = f_t(\text{label}_y)$. Given a test payload $p$, the predicted type is:
\[
\hat{y} = \argmin_{y \in \{1, \ldots, 15\}} \text{dist}(f_p(p), \bar{e}_y),
\]
where $\text{dist}(\cdot,\cdot)$ is cosine distance between L2-normalized embeddings. We use generic labels such as ``SQL injection attack'' rather than labels present in a training set to avoid shortcut learning -- the model has to generalize to abstract class descriptions not seen during training.
For more information on
this, see Appendix \ref{ap:labels}.

This design opens the possibility of zero-shot inference: a new vulnerability type could be introduced by specifying only a textual description, potentially enabling classification of payloads from previously unseen classes without retraining.

\subsection{Training Details}

In Stage 1, we train the model using Cached Multiple Negatives Ranking Loss~\cite{2021scaling} with an effective batch size of 4096, achieved through gradient accumulation. This large batch size increases the diversity of negative samples, providing a stronger contrastive learning signal.
In Stage 2, we switch to Mean Squared Error (MSE) loss with a smaller batch size of 32. This reduction is necessary because encoding raw payloads -- up to 16KB in size -- is computationally expensive and limits how large the batch can be.
Both stages use AdamW optimization with learning rate $4 \times 10^{-4}$ and early stopping to prevent overfitting. Embeddings are L2-normalized before computing distances.

\section{Evaluation Methodology}
\label{sec:methodology}

We evaluate \tool on threat classification: given an HTTP request-response pair, predict which of the 15 given vulnerability types the attacker attempted to exploit. We use two datasets: a large-scale private corpus with temporal splits, and a synthetic benchmark for controlled out-of-distribution testing. We compare against three baseline families: a BoW approach, off-the-shelf embeddings, and a classical fine-tuning method, which we consider susceptible to shortcut learning.

\subsection{Task and Data}
\label{sec:task}

Given an HTTP 1.1 request-response pair $p$, the model predicts vulnerability category $y \in \{1, \ldots, 15\}$. Both vulnerability descriptions and payloads originate from a proprietary threat intelligence database from Huawei. The database catalogs vulnerabilities with textual descriptions, metadata (CVE/CWE IDs, severity, publication dates), and labels.\footnote{https://isecurity.huawei.com/security/service/intelligence} Payloads are captured from production Intrusion Prevention Systems deployed across customer networks. More details are reported in Appendix \ref{ap:task_data}.

The 15 classes cover web exploitation techniques (XSS, Dir-traversal, Injection), attack outcomes (Code-execution, Info-Disclosure), and malware types (Trojan, Webshell, Worm, Botnet), among others. The taxonomy covers different abstraction levels and reflects the vendor's operational requirements rather than strict CWE alignment. Some classes semantically overlap (e.g., Injection vs Code-execution), complicating pattern learning from descriptions. All samples contain HTTP payloads, yet several classes represent vulnerabilities not directly associated with web applications (e.g., Overflow). In these cases, security analysts identified that the payload targets a specific product vulnerability using an HTTP-based attack vector. We use the original labeling as provided, testing \tool's ability to learn these operational patterns from textual vulnerability descriptions. We acknowledge these class definitions as a limitation, but they represent a realistic benchmark since models must learn to reproduce labeling derived from imperfect human-guided classification.

\textbf{Private operational dataset.}
The private dataset contains 29,675 textual vulnerability descriptions and 601,518 payloads across the 15 classes. The data is highly imbalanced (e.g., Code-execution: 167\,k payloads; Trojan: 72).
We partition by publication date on January~1, 2023: 25,526 threats (517,692 payloads) for training, 434 threats (83,826 payloads) for testing. This temporal split simulates zero-day conditions where the model must generalize to novel techniques disclosed after training. Descriptions average 127 tokens (range: 15--512); payloads average 1,847 bytes (range: 128--16,384). Multiple threat instances may share the same description when they represent minor variants of the same vulnerability.
From the training split, we extract 127,630 description triplets for Stage~1 and 517,692 text-payload pairs for Stage~2.

\textbf{Synthetic dataset.}
To enable reproducible evaluation, we construct a synthetic benchmark from public Indicators of Compromise (IoCs) sourced from the \textit{PayloadsAllTheThings} repository.\footnote{\url{https://github.com/swisskyrepo/PayloadsAllTheThings}} We prompt three LLMs (Gemini 2.5 Flash, DeepSeek-V3, Qwen3-Max) to generate realistic HTTP request-response pairs for each vulnerability type, using IoCs as guidance. The prompting template instructs models to produce complete HTTP transactions with varied headers, parameters, and server responses. 
Appendix \ref{ap:synthetic} details the generation
procedure.

The final dataset contains 11,000 balanced samples across 11 of the 15 classes. Four categories from the private dataset are excluded because public IoCs provide only high-level indicators (IP addresses, hostnames), insufficient to guide realistic payload generation. The vendor's internal labeling practices further complicate synthetic data generation: general-purpose LLMs are not aware of the vendor's internal standards and cannot precisely distinguish all classes without further instruction. Stage~1 of our pipeline is designed to address this challenge by aligning open LLM knowledge to task-specific practices. Despite these limitations, the synthetic benchmark provides controlled distributional shift: payloads are generated independently of the private training data, enabling reproducible testing of generalization to novel attack patterns.

More details about the datasets are reported in Appendix \ref{ap:task_data}.

\subsection{Baselines}
\label{sec:baselines}

Training uses only real data. Evaluation uses labeled payloads from both private (83\,k) and synthetic (11\,k) test sets. Classification for \tool proceeds via semantic retrieval (Section~\ref{sec:method}): embed the payload with $f_p$, compute cosine distance to centroids derived from $f_t$, and predict the nearest type. 

We compare against three baselines spanning distinct modeling paradigms. Appendix \ref{ap:baseline} provides full
implementation details.

The first baseline, \textit{TF-IDF + Random Forest (TF-IDF+RF)}, represents payloads as sparse TF-IDF vectors and trains a Random Forest classifier. This lexical approach reveals the extent to which token frequencies alone discriminate vulnerability types.
The second baseline, \textit{Fine-tuned CodeBERT + MLP (FT CodeBERT+MLP)}, fine-tunes CodeBERT on the classification task, extracting the \texttt{[CLS]} token representation and passing it to a multi-layer perceptron. This supervised baseline provides direct architectural comparison to our contrastive pretraining approach.
The third baseline, \textit{Embedding Similarity (RAG-style)}, embeds payloads with Sentence-BERT (all-MiniLM-L6-v2), builds an HNSW index of training samples, and classifies via k-nearest-neighbor majority voting. This retrieval baseline tests whether nearest-neighbor search in a generic embedding space suffices without contrastive refinement.
Together, these baselines establish reference points for lexical, parametric, and retrieval-based approaches, enabling assessment of the specific contributions of multi-modal contrastive alignment.

\section{Results}
\label{sec:results}

\begin{table}[!t]
\centering
\caption{Performance (\%) on real and synthetic sets. Relative improvements computed against best baseline per metric.}
\resizebox{.48\textwidth}{!}{
\begin{tabular}{lcccc}
\toprule
\multirow{2}{*}{\textbf{Method}} &
\multicolumn{2}{c}{\textbf{Real}} &
\multicolumn{2}{c}{\textbf{Synthetic}} \\
& \textbf{Acc.} & \textbf{Macro F1} & \textbf{Acc.} & \textbf{Macro F1} \\
\midrule
TF-IDF + RF & 65.7 & 29.8 & 16.0 & 9.0 \\
FT CodeBERT + MLP & 62.3 & 26.0 & 20.6 & 13.2 \\
Embedding Similarity & 48.9 & 20.9 & 8.1 & 4.3 \\
\tool & \textbf{68.1} & \textbf{30.1} & \textbf{24.4} & \textbf{14.3} \\
\midrule
Relative Improvement & +4\% & +1\% & +19\% & +8\% \\
\bottomrule
\end{tabular}}
\label{tab:results_private_and_synthetic}
\end{table}

\subsection{Quantitative Classification Results}
\label{sec:main_results}

Table~\ref{tab:results_private_and_synthetic} summarizes performance on both test sets. We report accuracy and macro F1.

On the real-world temporal split, \tool achieves 68.1\% accuracy, outperforming TF-IDF+RF (65.7\%, +2.4\%) and FT CodeBERT+MLP (62.3\%, +5.8\%). This indicates that contrastive learning with modal alignment yields more discriminative representations than both lexical methods and supervised fine-tuning. The macro F1 improvement (+0.3\% over TF-IDF+RF) is very low, reflecting persistent difficulty with less frequent classes and classes where vendor documentation lacks clarity.

The supervised CodeBERT baseline performs worse than TF-IDF+RF, suggesting that end-to-end fine-tuning on highly imbalanced and temporally shifted data leads to overfitting, while TF-IDF provides more robustness under these conditions. The Embedding Similarity baseline (48.9\%) confirms that nearest-neighbor retrieval in a generic embedding space does not capture the task-specific semantic distinctions required for accurate classification.

On the synthetic benchmark, all models show reduced performance. \tool maintains an advantage (24.4\%) over the best baseline (FT CodeBERT+MLP: 20.6\%). The drop from 68\% on real data to 24\% on synthetic data reflects two factors: (i)~synthetic payloads cover only 11 of the 15 classes, limiting comparability, and (ii)~LLM-generated payloads introduce patterns not present in operational traffic. This distribution shift is useful for probing generalization to unseen structures. 

An interesting pattern across all methods is the gap between accuracy and macro F1. While \tool achieves 68.1\% accuracy, its macro F1 is only 30.1\%. This discrepancy arises because accuracy is dominated by high-frequency classes (Code-execution, Injection, Info-Disclosure), where all models perform reasonably well, while macro F1 equally weights all 15 classes -- including those with very few test samples where all methods fail. This highlights that the challenge is not only temporal shift but also class imbalance, which the vendor's operational dataset exhibits.

Overall, these results point again to the credibility crisis introduced in Section~\ref{sec:intro}: models trained on limited proprietary datasets learn superficial correlations that fail under realistic temporal shifts. Contrastive learning with modal alignment appears to only partially mitigate this by grounding representations in semantic vulnerability descriptions. The improvements across both realistic and synthetic distributions suggest the approach may reduce shortcut learning. These results call for further extensions with additional textual anchors that could improve transferability and absolute performance.

\subsection{Qualitative Evaluation of Representations}
\label{sec:representation_quality}

\begin{figure}[!t]
    \centering
    \begin{subfigure}[b]{0.48\columnwidth}
        \includegraphics[width=\linewidth]{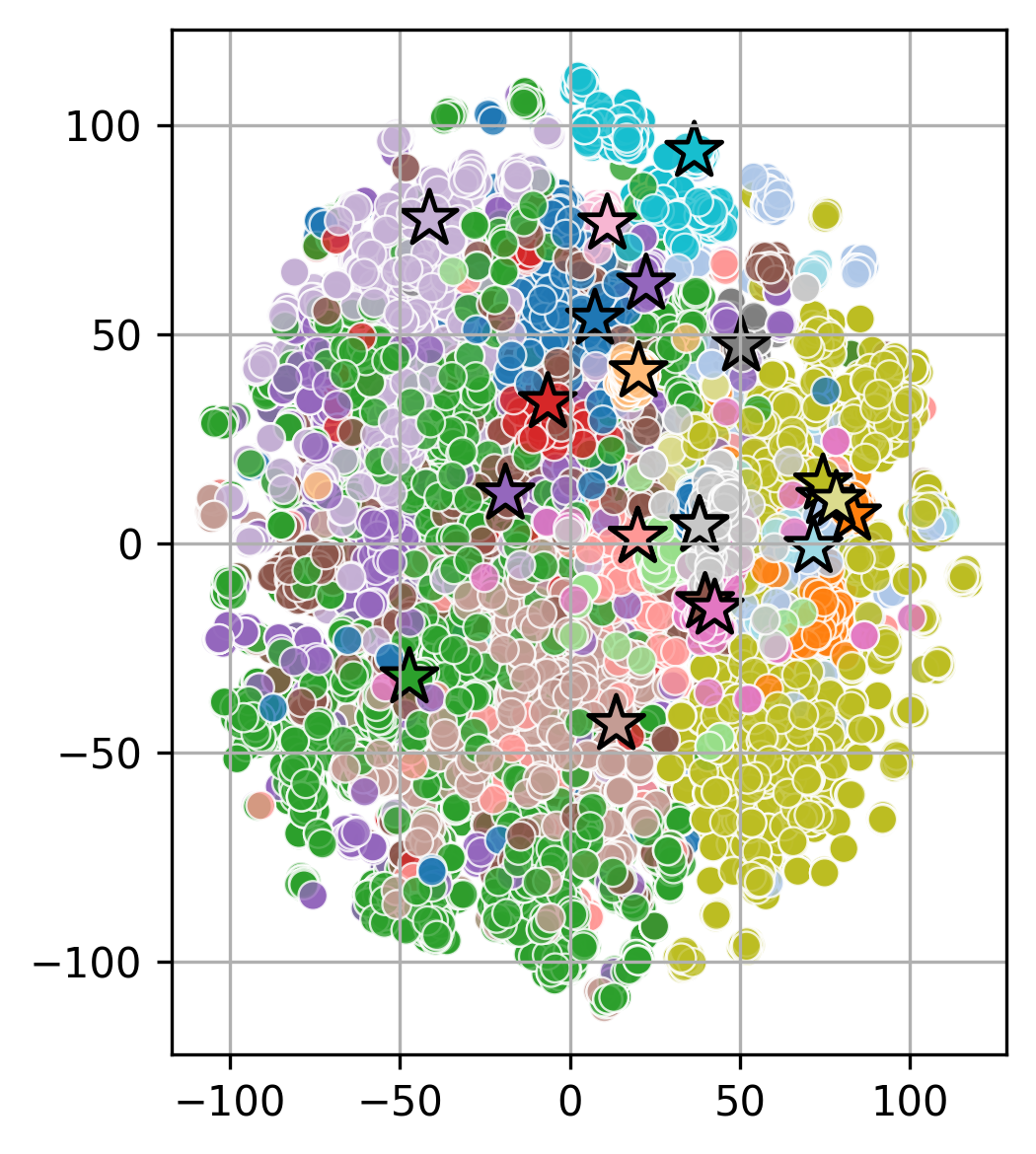}
        \caption{Original Model}
    \end{subfigure}
    \hfill
    \begin{subfigure}[b]{0.48\columnwidth}
        \includegraphics[width=\linewidth]{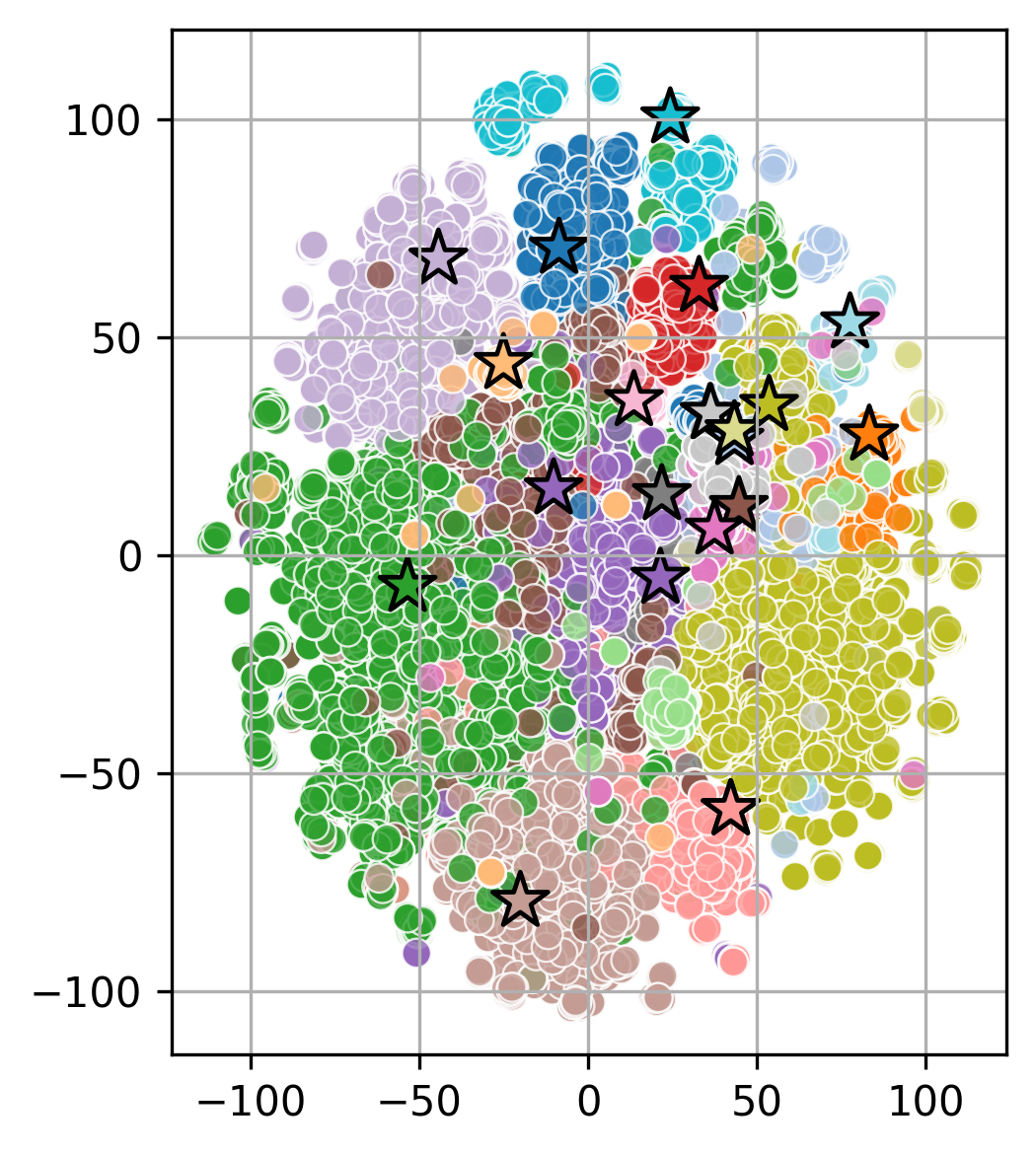}
        \caption{After Stage 1}
    \end{subfigure}
    \caption{t-SNE projections of vulnerability-description embeddings.
    Dots represent individual descriptions; Stars mark the generic anchor for each vulnerability type.
    Stage~1 contrastive learning structures the space around these anchors.
    }
    \label{fig:tsne_descriptions}
\end{figure}

\begin{figure}[!t]
    \centering
    \begin{subfigure}[b]{0.32\columnwidth}
        \includegraphics[width=\linewidth]{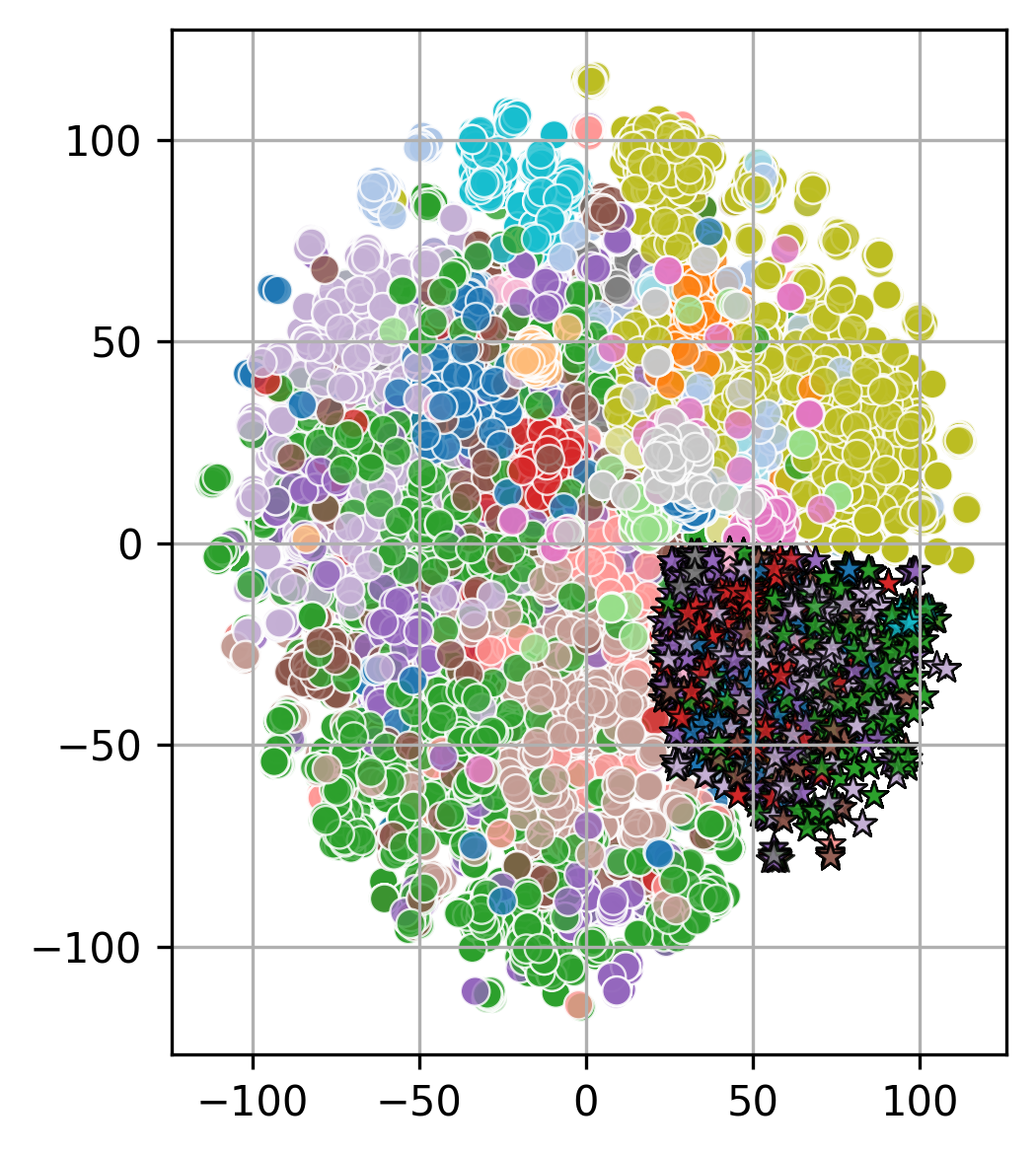}
        \caption{Original}
    \end{subfigure}
    \hfill
    \begin{subfigure}[b]{0.32\columnwidth}
        \includegraphics[width=\linewidth]{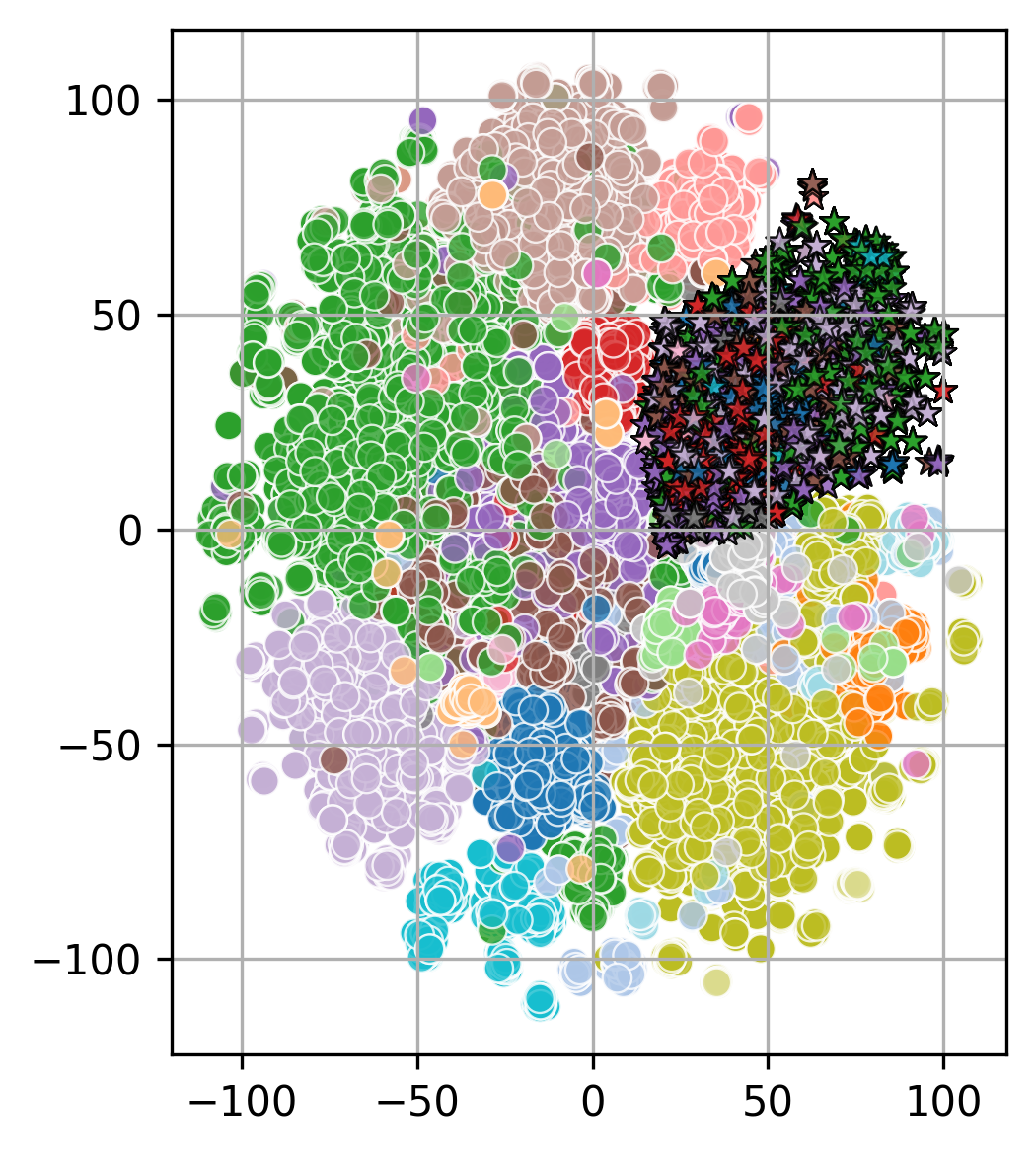}
        \caption{After Stage 1}
    \end{subfigure}
    \hfill
    \begin{subfigure}[b]{0.32\columnwidth}
        \includegraphics[width=\linewidth]{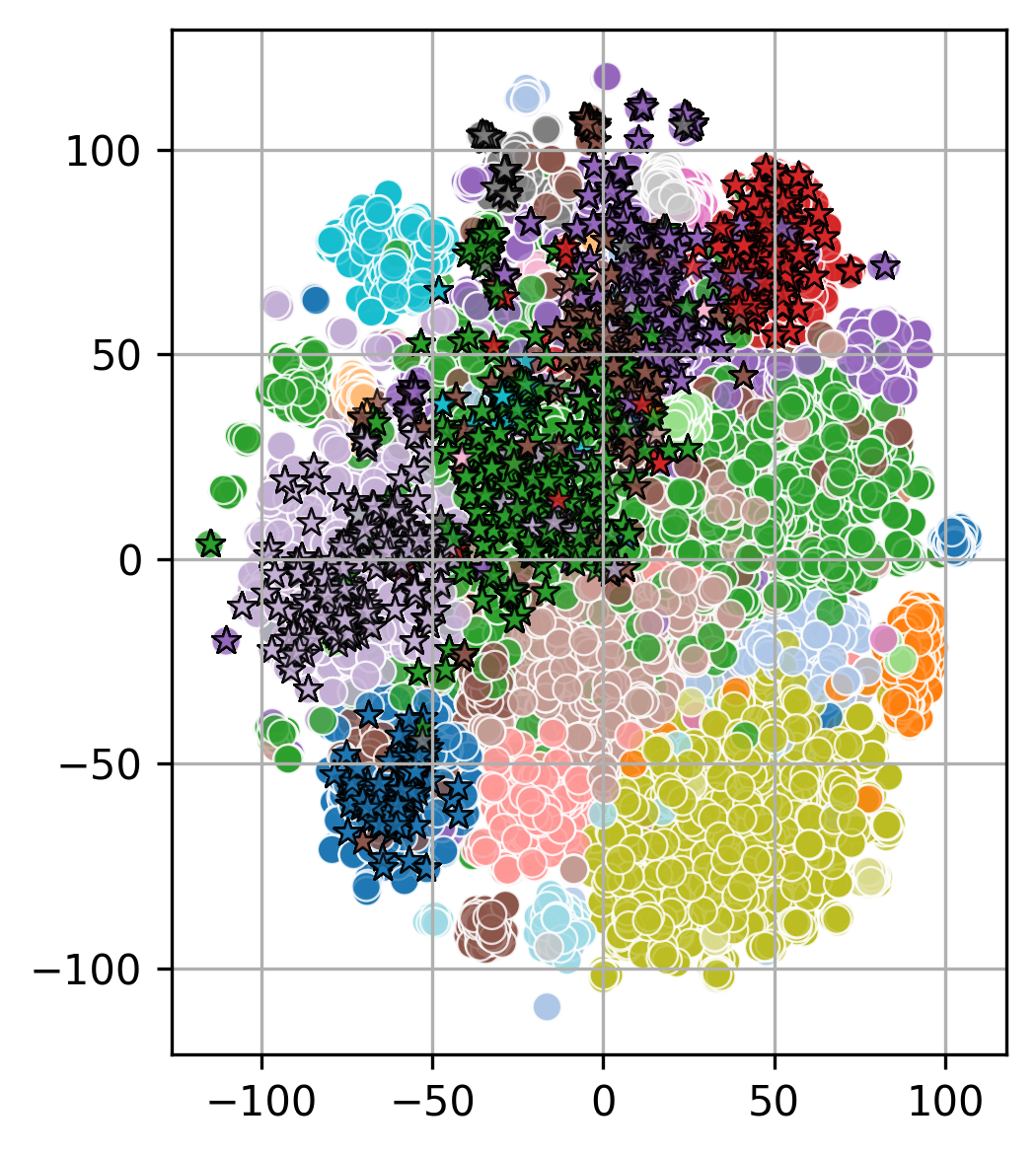}
        \caption{After Stage 2}
            \label{fig:tsne_payloads_c}

    \end{subfigure}
    \caption{t-SNE projections of payload embeddings across training stages (dots are descriptions, stars are payloads). Semantic structure emerges only after Stage~2.}
    \label{fig:tsne_payloads}
\end{figure}

We analyze how the representation space evolves throughout training to illustrate the impact of each stage. Figure~\ref{fig:tsne_descriptions} shows t-SNE projections of vulnerability-description embeddings at two points: the pretrained model and after Stage~1 contrastive learning. Each dot corresponds to an individual vulnerability-description instance from the dataset, while each star represents the generic textual anchor for a vulnerability type used during semantic retrieval. In the pretrained model, descriptions are highly entangled -- the generic pretrained model is not precisely aligned to the vulnerability descriptions of this specific cybersecurity task. The vendor's descriptions are roughly grouped, but the class definitions are not immediately evident in the embedding space. After Stage~1, the contrastive objective produces coherent clusters centered around their generic type descriptions. Because the text encoder is frozen in Stage~2, this structure is preserved.

Figure~\ref{fig:tsne_payloads} shows how payload embeddings evolve across all three stages. Initially, payloads form an unstructured mass with no clustering by vulnerability type. They are also anomalies with respect to the embedding space built with descriptions -- expected since the initial model is pretrained with text and has no knowledge about payload samples. Since Stage~1 trains exclusively on textual descriptions without involving payloads, their embeddings remain unorganized after this stage, even though the textual embeddings are better organized. The transformation occurs in Stage~2: once the payload encoder is trained to match the frozen description embeddings through modal alignment, it learns the same semantic structure as the text space. The final projection (Figure~\ref{fig:tsne_payloads_c}) shows coherent clusters by vulnerability type, mirroring the organization established in the text space.

These visualizations suggest that the two-stage framework produces a more structured space organized by vulnerability type. Stage~1 establishes semantic structure in text; Stage~2 transfers this structure to payloads. The final result is a shared embedding space where both modalities cluster around vulnerability-type prototypes defined by textual anchors.

\subsection{Per-Class Observations}

\begin{table}[!t]
\centering
\caption{Per-class F1-scores (\%) on the challenging temporal split. Only classes present in this partition are shown. A dash (--) indicates zero F1. Best results per class in bold.}
\label{tab:per_class_f1}
\resizebox{.48\textwidth}{!}{
\begin{tabular}{lrrrr}
\toprule
\textbf{Class} & \textbf{FT CB} & \textbf{Embed.} & \textbf{TF-IDF} & \textbf{\tool} \\
\midrule
Code-execution   & 51.5 & 43.9 & \textbf{60.2} & 50.3 \\
Dir-traversal    & 63.0 & 46.6 & 60.1 & \textbf{78.6} \\
DoS              & --    & --    & --    & -- \\
Info-Disclosure  & 57.3 & 51.3 & 63.0 & \textbf{70.8} \\
Injection        & \textbf{73.4} & 55.7 & 69.8 & 69.2 \\
RFI              & --    & --    & --    & -- \\
Scanner          & 15.3 & 15.3 & 4.0 & 6.1 \\
Trojan           & --    & --    & --    & -- \\
Webshell         & 9.3 & 13.9 & 6.7 & 0.3 \\
XSS              & \textbf{93.7} & 66.2 & 94.0 & 85.7 \\
\bottomrule
\end{tabular}}
\end{table}

Table~\ref{tab:per_class_f1} shows per-class F1-scores on the challenging temporal split. \tool provides the strongest improvements for classes with clear structural patterns: \textit{Dir-traversal} (F1: 78.6 vs.\ 63.0 for the best baseline) and \textit{Info-Disclosure} (70.8 vs.\ 63.0). These classes retain usable signal under temporal shift because their payloads exhibit consistent syntactic behaviors that contrastive training encodes robustly. In contrast, rare or ambiguous classes such as \textit{Trojan}, \textit{Webshell}, and \textit{Remote-File-Inclusion} remain difficult for all methods, with near-zero F1 scores due to data scarcity and semantic overlap in the vendor's taxonomy. Improvements correlate with classes whose textual semantics are well captured during Stage~1, suggesting that richer vulnerability descriptions could further improve performance.

In the easy split (random stratified, not shown), almost all classes achieve high F1 under all methods, reflecting homogeneous training-test conditions. The challenging temporal split reveals the true picture: many classes collapse to zero F1 for all methods, while \tool maintains performance where the textual modality provides clear discriminative semantics. Notably, for \textit{Code-execution} and \textit{Injection} -- the two largest classes -- TF-IDF+RF remains competitive or superior, likely because token-level features are sufficiently stable for these well-represented categories. \tool's advantage concentrates on mid-frequency classes where lexical cues are less reliable but textual descriptions provide clear semantic grounding.

\section{Limitations and Conclusion}
\label{sec:conclusion}

We studied contrastive multi-modal learning as a means to mitigate shortcut learning in cybersecurity tasks. We introduced \tool, a two-stage framework that grounds payload representations in textual vulnerability descriptions. Our framework structures text space via contrastive learning, then aligns payloads to this frozen space. It achieves 0.68 accuracy on temporally shifted private data and appears to keep some advantages on synthetic out-of-distribution tests, with the potential to reduce dependence on large annotated payload corpora by transferring semantic structure from text to the payload modality.

Several limitations remain. First, we evaluate on a single cybersecurity task; while we find consistent improvements over baselines across temporal splits and synthetic distributions, broader validation on other tasks is needed before claiming general applicability. Second, 0.68 accuracy remains far from production-grade reliability. Our intuition is that several factors prevent better results: the vendor's classes mix abstraction levels (exploitation techniques, attack outcomes, and vectors), textual descriptions lack detail for some classes, and the dataset exhibits strong class imbalance. In other words, achieving better results requires better data -- yet we believe our approach will reduce the need for large datasets, a question we plan to investigate. Third, the temporal split tests generalization to novel techniques within seen types, not zero-shot transfer to entirely new categories. While \tool naturally supports such transfer, properly validating that capability requires dedicated experiments.

The improvements across realistic and synthetic distributions suggest the method captures more robust representations than purely supervised approaches. To help others explore the methodology, we release our synthetic benchmark and source code. Promising directions include extending the approach to additional cybersecurity tasks (e.g., malware family classification, intrusion detection), incorporating richer textual sources, investigating more sophisticated triplet mining strategies to improve contrastive learning efficiency, and exploring whether the zero-shot transfer capability of \tool can be validated on new vulnerability categories.

\section*{Acknowledgments}
We thank Dr. Zied Ben Houidi for his contribution to the ideas that shaped this work and his insightful comments on the manuscript.

This work was supported by Huawei Technologies France under the project ``AISN -- AI Secured Networks'' and in part by the AI4CTI FISA under Project FISA-2023-00168, funded by the Italian Ministry of University and Research (MUR). 

\bibliographystyle{IEEEtran}
\bibliography{references}

\appendices

\section{Textual Anchors for Semantic Prototypes}
\label{ap:labels}

During inference, the model does not rely on any payload-derived statistics or classifier parameters. Instead, prediction is performed through semantic retrieval: a payload embedding is compared against a set of textual prototypes, one for each vulnerability type. These prototypes are encoded using short textual labels that summarize the core semantics of each class. The full collection of labels used in this work is provided in Table~\ref{tab:label_dict}.
The labels were designed with three goals:

\begin{enumerate}
\item \textbf{Generic, high-level descriptions:} Unlike the detailed descriptions found in the real dataset, which often contain specific exploit strings, versions, or product names, the prototype labels intentionally abstract away concrete details. Each label focuses on the underlying vulnerability mechanism (for example, ``arbitrary code execution'' or ``unauthorized file access''), along with brief explanations of impact and common mitigation strategies. This prevents the centroid from encoding dataset-specific artifacts and guides the model to retrieve based on categories.

\item \textbf{Single-description centroids for zero-shot evaluation:}
Instead of computing centroids as the mean of many description embeddings, we use a single label per vulnerability type. This design tests the model’s ability to align payloads with a textual representation it has \emph{not} seen during training. Because the prototype is generic, the retrieval mechanism can be evaluated in zero-shot conditions. If a new vulnerability type emerges, a prototype label can be added without retraining.
\end{enumerate}

\begin{table*}[t]
\centering
\caption{Textual vulnerability-type prototypes used as anchors during semantic retrieval. Each description provides a concise characterization of the vulnerability type for the contrastive learning stage.}
\label{tab:label_dict}
\small
\begin{tabular}{@{}lp{5cm}lp{5cm}@{}}
\toprule
\textbf{Vulnerability Type} & \textbf{Prototype Description} & \textbf{Vulnerability Type} & \textbf{Prototype Description} \\
\midrule
Backdoor & A hidden entry point allowing unauthorized system access. Data breaches and loss of control. & 
Injection & A vulnerability that lets attackers inject malicious code. Unauthorized database manipulation and system compromise. \\
\midrule
Botnet & A network of compromised computers controlled remotely. Large-scale cyber-attacks and unauthorized system exploitation. & 
Overflow & A memory-related flaw causing crashes or unauthorized access. Unexpected execution paths and system instability. \\
\midrule
CGI & A flaw in web applications enabling unauthorized actions. Data theft, code execution, and service disruption. & 
Remote-File-Inclusion & A vulnerability permitting attackers to include external malicious files. Unauthorized remote file execution and system compromise. \\
\midrule
Code-Execution & A security flaw allowing arbitrary code execution. Unauthorized system control and potential data breaches. & 
Scanner & Automated tools probing for system vulnerabilities. Potential exposure of security gaps exploitable by attackers. \\
\midrule
Dir-Traversal & A security issue enabling attackers to retrieve sensitive files. Unauthorized file access and system compromise. & 
Trojan & Malware disguised as legitimate software for system compromise. Unauthorized access and data theft. \\
\midrule
DoS & An attack aimed at rendering a system unusable. Network disruption and resource exhaustion. & 
Webshell & A hidden backdoor allowing remote system control. Data manipulation and complete system takeover. \\
\midrule
Info-Disclosure & A weakness that allows unauthorized access to internal data. May lead to identity theft or system mapping. & 
Worm & A self-propagating malware spreading without human intervention. Network-wide infection and resource depletion. \\
\midrule
& & 
XSS & A vulnerability allowing unauthorized script execution in web applications. Data theft and session hijacking. \\
\bottomrule
\end{tabular}
\end{table*}

\section{Classes and dataset distributions}
\label{ap:task_data}

The 15-class problem reflects requirements defined by the vendor based on pattern matching and manual security analyst review. Table~\ref{tab:threat_types} summarizes the categories with approximate CWE mappings.

\begin{table*}[!t]
\centering
\caption{The 15 vulnerability classes used for threat classification. CWE mappings are approximate as the taxonomy reflects operational requirements rather than strict CWE alignment.}
\label{tab:threat_types}
\small
\begin{tabular}{@{}lp{11cm}l@{}}
\toprule
\textbf{Class} & \textbf{Description} & \textbf{Example CWE} \\
\midrule
Backdoor & Traffic associated with hidden methods to gain unauthorized remote access & CWE-912 \\
Botnet & Traffic from compromised systems controlled by attackers for coordinated attacks & N/A \\
CGI & Payload targeting CGI script vulnerabilities due to improper input validation & CWE-77, CWE-20 \\
Code-execution & Payload carrying attempts to execute arbitrary commands or code on the target system & CWE-94 \\
Dir-traversal & Path traversal payload attempting to access files outside the web root directory & CWE-22 \\
DoS & Denial of Service traffic attempting to exhaust resources and make them unavailable & CWE-400 \\
Info-Disclosure & Traffic carrying unauthorized exposure of sensitive information to attackers & CWE-200 \\
Injection & Payload containing command, SQL or other injections (e.g., SQL, Command) & CWE-74, CWE-89 \\
Overflow & Payload attempting to exploit a known buffer overflow vulnerability in a product & CWE-119 \\
Remote-File-Inclusion & Payload attempting to include and execute remote files via URL & CWE-98 \\
Scanner & Traffic from automated vulnerability scanning tools probing for weaknesses & N/A \\
Trojan & Traffic associated with a general group of malware software & CWE-507 \\
Webshell & Traffic from malicious script uploaded to web server enabling remote administration & CWE-434 \\
Worm & Traffic from another specific malware group spreading through network vulnerabilities & N/A  \\
XSS & Payload injecting malicious scripts into trusted websites (Cross-Site Scripting) & CWE-79 \\
\bottomrule
\end{tabular}
\end{table*}

Table~\ref{tab:unified_support_table_reordered_final} shows the per-class distribution across all datasets. Note that these categories reflect the vendor's operational practices rather than strict alignment with open taxonomies. Indeed, some of the classes are clearly defined, whereas others semantically overlap, such as Trojan, Worm and Webshell. This overlap is identifiable in the text description of payloads belonging to the classes, which are often repetitive, and complicates the learning of generic patterns from the data.

\begin{table*}[!t]
\centering
\caption{Dataset statistics per class. Payloads in the operational dataset reflect labels provided by the operator's deployment; synthetic payloads are balanced for controlled evaluation. Descriptions drive Stage~1 contrastive learning.}
\label{tab:unified_support_table_reordered_final}
\begin{tabular}{lrrr}
\toprule
\textbf{Vulnerability Type} &
\textbf{Payloads (Real)} &
\textbf{Payloads (Synthetic)} &
\textbf{Descriptions} \\
\midrule
Backdoor              & 586      & 1,000 & 946   \\
Botnet                & 978      & 0     & 676   \\
CGI                   & 192      & 1,000 & 9     \\
Code-execution        & 166,359  & 0     & 7,805 \\
Dir-traversal         & 90,636   & 1,000 & 811   \\
DoS                   & 610      & 1,000 & 1,164 \\
Info-Disclosure       & 113,851  & 1,000 & 2,025 \\
Injection             & 143,112  & 1,000 & 3,090 \\
Overflow              & 718      & 1,000 & 3,291 \\
Remote-File-Inclusion & 5,515    & 1,000 & 122   \\
Scanner               & 14,654   & 0     & 419   \\
Trojan                & 72       & 1,000 & 6,256 \\
Webshell              & 10,684   & 0     & 1,221 \\
Worm                  & 290      & 1,000 & 374   \\
XSS                   & 53,261   & 1,000 & 1,466 \\
\midrule
\textbf{Total}        & 601,518 & 11,000 & 29,675 \\
\bottomrule
\end{tabular}
\end{table*}
\section{Synthetic Dataset}
\label{ap:synthetic}

To complement evaluation on the private operational dataset, we construct a balanced synthetic dataset designed to provide a controlled out-of-distribution benchmark. Starting from publicly available Indicators of Compromise (IoCs) and payload fragments collected from open repositories, we use them as guidance signals to drive the generation of realistic HTTP request–response pairs.

Using these IoCs and fragments, we prompt three large language models, Gemini (Gemini~2.5~Flash), DeepSeek (DeepSeek-V3), and Qwen (Qwen3-Max), to synthesize malicious HTTP traffic for each vulnerability category. The models produce complete attack samples, each consisting of a full request and a plausible server-side response, while varying structural elements such as headers, query parameters, injected payloads, and contextual behaviors. A simplified version of the prompting template is shown in Listing~\ref{lst:synthetic_prompt}.

\begin{lstlisting}[style=coloredlisting, caption={Prompt template used to generate synthetic attack samples.}, label={lst:synthetic_prompt}]
You are given files containing file names related to a specific vulnerability type Worm.

**Strictly use only the information and examples derived from those files**. Do NOT invent payloads or behaviors beyond what is described.

Generate **250 realistic and complete HTTP payload samples for education** that simulate typical traffic of **successful attacks** based on the provided files.

### Requirements:
1. Each sample must include a **complete and valid HTTP request** (GET or POST) and a **complete HTTP response**, both written as raw HTTP text.
   - The request must contain **at least 8 realistic HTTP headers**, and their values must vary across samples.
   - Include plausible query parameters or POST body fields relevant to the attack scenario.
2. The **response** must reflect a realistic server behavior after a successful attack (e.g., returning sensitive data, executing attacker logic, authentication bypass, internal errors, injected content).
3. Output format: A **valid JSON array** of objects. Each object must include:
   - `Category`: The attack category.
   - `HTTP Payload`: The raw HTTP request and response separated by `\n\n---RESPONSE---\n\n`.
4. Samples must be **diverse and unique**, covering different patterns and techniques discussed in the files.
5. The output must be **valid JSON with no extra text**.

### Example Output:
[
  {
    "Category": "Injection",
    "HTTP Payload": "GET /vulnerable.php?cmd=cat%20/etc/passwd HTTP/1.1\\r\\nHost: vulnerable.site ... (truncated)"
  }
]
\end{lstlisting}

Repeated prompting with different seeds and model configurations yields substantial variation in structure, headers, payload encoding styles, and server responses. In total, the process produces 11\,000 synthetic samples across 11 vulnerability categories. Some classes could not be generated reliably due to insufficient IoCs or payload fragments, particularly categories where public evidence consists only of high-level indicators such as IP addresses rather than concrete HTTP exploit traces. This synthetic dataset is used exclusively for evaluation.

\section{Baseline Models}
\label{ap:baseline}

This appendix provides the detailed descriptions of the three baseline models used for comparison throughout the paper.

\subsection{Traditional TF-IDF + Random Forest (TF-IDF+RF)}

This classical machine learning pipeline transforms raw payload text into sparse Term Frequency–Inverse Document Frequency (TF-IDF) vectors, which capture word-importance patterns without modeling contextual semantics. A Random Forest classifier is then trained on these features. Despite its simplicity, this baseline is interpretable and highlights the extent to which lexical cues alone can discriminate between vulnerability types.

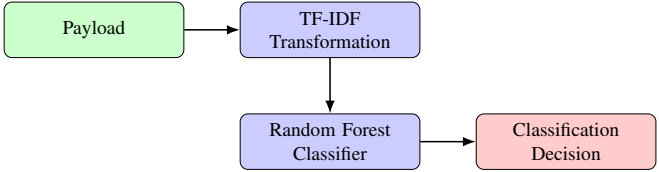
\begin{figure}[htb]
\centering
\adjustbox{width=\linewidth}{
\begin{tikzpicture}[
node distance=1cm,
block/.style={rectangle, draw, fill=blue!20, text width=3cm, text centered, rounded corners, minimum height=1cm},
input/.style={rectangle, draw, fill=green!20, text width=3cm, text centered, rounded corners, minimum height=1cm},
output/.style={rectangle, draw, fill=red!20, text width=3cm, text centered, rounded corners, minimum height=1cm},
arrow/.style={-Latex, thick}
]
\node (input) [input] {Payload};
\node (tfidf) [block, right=of input] {TF-IDF Transformation};
\node (rf) [block, below=of tfidf] {Random Forest Classifier};
\node (output) [output, right=of rf] {Classification Decision};

\draw [arrow] (input) -- (tfidf);
\draw [arrow] (tfidf) -- (rf);
\draw [arrow] (rf) -- (output);
\end{tikzpicture}
}
\caption{Traditional TF-IDF + Random Forest pipeline.}
\label{fig:tfidf_rf_pipeline}
\end{figure}

\subsection{Fine-tuned CodeBERT Classifier (FT CodeBERT+MLP)}
\label{ap:codebert}

This baseline fine-tunes the pretrained CodeBERT encoder directly on the classification task. The contextual representation of the \texttt{[CLS]} token is extracted and fed into a small multilayer perceptron (MLP) to predict the vulnerability type. This setup mirrors our model’s architecture but lacks contrastive pretraining and modal alignment, allowing us to isolate the contribution of our training strategy.

\begin{figure}[htb]
\centering
\adjustbox{width=\linewidth}{
\begin{tikzpicture}[
node distance=1cm,
block/.style={rectangle, draw, fill=blue!20, text width=3cm, text centered, rounded corners, minimum height=1.2cm},
transformer/.style={rectangle, draw, fill=purple!20, text width=3cm, text centered, rounded corners, minimum height=1.5cm, line width=1pt},
input/.style={rectangle, draw, fill=green!20, text width=3cm, text centered, rounded corners, minimum height=1cm},
output/.style={rectangle, draw, fill=red!20, text width=3cm, text centered, rounded corners, minimum height=1cm},
arrow/.style={-Latex, thick},
token/.style={circle, draw, fill=orange!20, minimum size=0.8cm, inner sep=0pt, font=\small},
mlp/.style={rectangle, draw, fill=cyan!20, text width=3cm, text centered, rounded corners, minimum height=1cm}
]

\node (input) [input] {Payload};
\node (codebert) [transformer, right=of input, xshift=2cm] {
\textbf{CodeBERT} \\
(Fine-tuned Transformer)
};

\node (cls_token) [token, below=of codebert] {[CLS]};
\node (mlp) [mlp, below=of cls_token] {MLP Classifier};
\node (output) [output, right=of mlp] {Prediction};

\draw [arrow] (input) -- (codebert);
\draw [arrow] (codebert.south) -- (cls_token);
\draw [arrow] (cls_token) -- (mlp);
\draw [arrow] (mlp) -- (output);

\end{tikzpicture}
}
\caption{Fine-tuned CodeBERT + MLP pipeline.}
\label{fig:codebert_mlp_pipeline}
\end{figure}
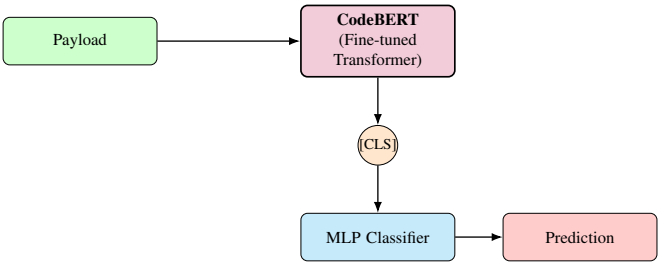

\subsection{Embedding Similarity Classifier (RAG-style Retrieval)}

Inspired by retrieval-augmented methods, this classifier embeds each payload using a fine-tuned Sentence-BERT encoder (all-MiniLM-L6-v2). During inference, the embedding is queried against a pre-built HNSW index of labeled samples. The top-$k$ nearest neighbors are retrieved, and their labels are combined via majority voting. This approach leverages similarity in embedding space rather than supervised fine-tuning.

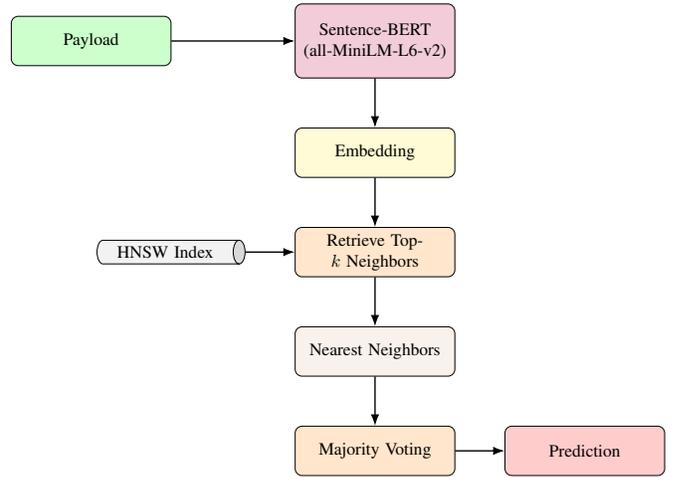
\begin{figure}[htb]
\centering
\adjustbox{width=\linewidth}{
\begin{tikzpicture}[
node distance=1cm,
block/.style={rectangle, draw, fill=blue!20, text width=3cm, text centered, rounded corners, minimum height=1cm},
input/.style={rectangle, draw, fill=green!20, text width=3cm, text centered, rounded corners, minimum height=1cm},
output/.style={rectangle, draw, fill=red!20, text width=3cm, text centered, rounded corners, minimum height=1cm},
model/.style={rectangle, draw, fill=purple!20, text width=3cm, text centered, rounded corners, minimum height=1.5cm},
index/.style={cylinder, cylinder uses custom fill, cylinder body fill=gray!10, cylinder end fill=gray!30, draw, text width=2.5cm, text centered, minimum height=2cm, aspect=0.5},
process/.style={rectangle, draw, fill=orange!20, text width=3cm, text centered, rounded corners, minimum height=1cm},
arrow/.style={-Latex, thick}
]

\node (input) [input] {Payload};
\node (sentence_bert) [model, right=of input, xshift=1.5cm] {
Sentence-BERT \\
(all-MiniLM-L6-v2)
};
\node (embedding) [block, below=of sentence_bert, fill=yellow!20] {Embedding};
\node (retrieval) [process, below=of embedding] {Retrieve Top-$k$ Neighbors};
\node (hnsw) [index, left=of retrieval] {HNSW Index};
\node (neighbors) [block, below=of retrieval, fill=brown!10] {Nearest Neighbors};
\node (voting) [process, below=of neighbors] {Majority Voting};
\node (output) [output, right=of voting] {Prediction};

\draw[arrow] (input) -- (sentence_bert);
\draw[arrow] (sentence_bert) -- (embedding);
\draw[arrow] (embedding) -- (retrieval);
\draw[arrow] (hnsw) -- (retrieval);
\draw[arrow] (retrieval) -- (neighbors);
\draw[arrow] (neighbors) -- (voting);
\draw[arrow] (voting) -- (output);

\end{tikzpicture}
}
\caption{Retrieval-augmented embedding similarity classifier.}
\label{fig:rag_pipeline}
\end{figure}

%


\end{document}